
\magnification=\magstep1\hsize=13cm\vsize=20cm\overfullrule 0pt
\baselineskip=13pt plus1pt minus1pt
\lineskip=3.5pt plus1pt minus1pt
\lineskiplimit=3.5pt
\parskip=4pt plus1pt minus4pt

\def\negenspace{\kern-1.1em}\def\quer{\negenspace\nearrow}



\def\ltextindent#1{\hbox to \hangindent{#1\hss}\ignorespaces}



\def\sqr#1#2{{\vcenter{\hrule height.#2pt\hbox{\vrule width.#2pt
height#1pt \kern#1pt \vrule width.#2pt}\hrule height.#2pt}}}

\newcount\refno
\refno=1
\def\y{\the\refno}
\def\myfoot#1{\footnote{$^{(\y)}$}{#1}
                 \advance\refno by 1}


\def\newref{\vskip 1pc 
            \hangindent=2pc
            \hangafter=1
            \noindent}

\def\neq{\hbox{$\,$=\kern-6.5pt /$\,$}}


\def\semidirect{\;{\rlap{$\subset$}\times}\;}



\newcount\secno
\secno=0
\newcount\fmno\def\z{\global\advance\fmno by 1 \the\secno.
                       \the\fmno}
\def\sectio#1{\medbreak\global\advance\secno by 1
                  \fmno=0
     \noindent{\the\secno. {\it #1}}\noindent}

\noindent{\it file avoid7.tex, 1992-12-30,
submitted to Phys.\ Rev.\ D}
\medskip
\centerline{\bf AVOIDING DEGENERATE COFRAMES}
 \centerline{\bf IN AN AFFINE GAUGE APPROACH TO QUANTUM GRAVITY}
\bigskip
\centerline{by}
\bigskip
\centerline{Eckehard W. Mielke$^{*\$ }$,
J. Dermott McCrea$^{**}$, Yuval Ne'eman$^{*** \diamond \$ +)}$,}
\centerline{and Friedrich W. Hehl$^{*\$}$}
\medskip
\noindent $^{*})$ Institute for Theoretical Physics, University of
Cologne, D(W)-5000 K\"oln 41, Germany

\noindent $^{**})$ Department of Mathematical Physics,
University College, Dublin 4, and Dublin Institute for
Advanced Studies, Dublin 4, Ireland

\noindent $^{***})$ Raymond and Beverley Sackler Faculty of Exact
Sciences, Tel-Aviv University, Tel-Aviv 69978, Israel 
\bigskip \centerline{\bf Abstract} \bigskip In quantum models of
gravity, it is surmized that configurations with degenerate coframes
could occur during topology change of the underlying spacetime
structure. However, the coframe is not the true Yang--Mills type gauge
field of the translations, since it lacks the inhomogeneous gradient
term in the gauge transformations.  By explicitly restoring this
``hidden" piece within the framework of the affine gauge approach to
gravity, one can avoid the metric or coframe degeneracy which would
otherwise interfere with the integrations within the path integral.
This is an important advantage for quantization.  \medskip \medskip
\noindent PACS numbers: 04.50.+h, 11.15.-q, 12.25.+e .  \vfill
\noindent $^\diamond)$ On leave from the Center for Particle Physics,
University of Texas, Austin, Texas 78712, USA

\medskip
\noindent $^{\$})$ Supported by the German-Israeli Foundation for
Scientific Research \& Development (GIF), Jerusalem and Munich,
project I-52-212.7/87.

\noindent $^{+})$Supported in part by DOE Grant DE-FG05-85-ER40200.

\eject

\sectio{\bf Introduction}

As it should be recognized by now, ``...gravity is that field
which corresponds to a gauge invariance
with respect to displacement transformations", as Feynman [1] has
put it. On a macroscopic scale, gravity is empirically rather
well described by Einstein's general relativity theory (GR) which
resides in a curved pseudo--Riemannian spacetime.

In a first order formalism, one introduces a {\it local frame} field
(or vielbein), $e_{\alpha} = e^{i}{}_{\alpha}\,\partial_{i}\, $
which is expanded via the tetrad coefficients $ e^{i}{}_{\alpha}$
in terms of the coordinate basis $\partial_{i}:=
\partial /\partial x^{i}$, together with the {\it coframe} field
or $1$--form basis $\vartheta^{\beta} = e_{j}{}^{\beta}\,dx^{j}\>,$
which is dual to the frame $e_\alpha$ with respect to the {\it
interior product}: $e_{\alpha}\rfloor\vartheta^{\beta} =
e^{i}{}_{\alpha}\,e_{i}{}^{\beta} = \delta_{\alpha}^{\beta}\>$.
In the Introduction the frame is chosen to be (pseudo-){\it orthonormal}.
Quite often, $\vartheta^{\alpha}$ is advocated as the translational gauge
potential, although it does {\it not} transform inhomogeneously
under local frame rotations, as is characteristic for a connection.

The Ein\-stein--Cartan Lagrangian [2] is given by
$$V_{\rm EC}=-{1\over{ 4\ell^{2}}}{\buildrel\circ\over R}{}^{\alpha\beta}
\wedge\vartheta^{\gamma}\wedge \vartheta^{\delta}\,
\eta_{\alpha\beta\gamma\delta}\, ,\eqno(\z)$$
where ${\buildrel\circ\over R}{}^{\alpha\beta}=R^{[\alpha\beta]}$ is the
curvature 2-form
associated with the Lorentz connection 1--forms
${\buildrel\circ\over \Gamma}{}^{\alpha\beta}=\Gamma^{[\alpha\beta]}$ and
$\eta_{\alpha\beta\gamma\delta} :=
\sqrt{|det\; o_{\alpha\beta}|}\;\epsilon_{\alpha\beta\gamma\delta}$
is the Levi--Civita tensor. Because of the orthonormality chosen, the
local metric components read $o_{\alpha\beta}=\hbox{diag}(1,-1,-1,-1)$.
In constructing macroscopic viable
gravitational Lagrangians, the fundamental length $\ell$ needs to
be identified with the Planck length $\ell_{\rm Planck}$.
Vacuum GR can be consistently
recovered by imposing the constraint of vanishing torsion
$T^{\alpha}:=D\vartheta^{\alpha}$ via the addition of the Lagrange
multiplier term $\mu_{\alpha}\wedge T^{\alpha}$ to (1.1). Then
only the Belinfante--Rosenfeld  symmetrized
energy--momentum current occurs as a source of gravity, the contribution
of the matter spin being subtracted out.

One avenue of quantizing gravity is to consider the functional integral
$$\int {\cal D}\vartheta\; {\cal D}{\buildrel\circ\over \Gamma}\;
exp\bigr( i\int_{M}V_{\rm EC}\bigl),\eqno(\z)$$
where a summation is understood over all {\it inequivalent} coframes
$\vartheta:=\vartheta^{\alpha}\,P_{\alpha}$, Lorentz connections
${\buildrel\circ\over \Gamma}:={\buildrel\circ\over \Gamma}{}^{\alpha\beta}
\,{\buildrel\circ\over L}_{\alpha\beta}\,$, and spacetime topologies, as
well. Since in GR the Lorentz connection
${\buildrel\circ\over \Gamma}{}^{\alpha\beta}$ is constrained by
$T^{\alpha}=0$, Eq.\ (1.1) will become the Hilbert second order
Lagrangian and an integration over all coframes is sufficient. In
any case, this summation will also involve {\it degenerate}
($det\, e_{j}{}^{\beta}=0$) or even vanishing coframes. This instance
would induce the breakdown of any length measurement performed by means of
the metric $g= o_{\alpha\beta}\,\vartheta^{\alpha}\otimes\vartheta^{\beta}$
and would signal the possible occurence of a topology change, as is argued
in the interesting paper of Horowitz [3]. This also gives a flavor
of some of the conceptual difficulties [4] not encountered in the
quantization of internal Yang--Mills theories on a {\it fixed} spacetime
background.

Degenerate coframes are not only restricted to the realm of quantum
gravity. In Ashtekar's reformulation of canonical GR, cf. [5],
the {\it ``triad density"}, i.e. more precisely, the tangential
$2$--form $\,^{\underline{*}}\underline{\vartheta}^{\alpha}$, is in
fact allowed to become degenerate as a classical solution of
Hamilton's equations. Moreover, in a first order formulation of
topological 3D gravity [6], we uncovered a ``dynamical symmetry" in which
coframe and ``Lorentz"--rotational connection (in three dimensions) become
related\myfoot{This is analogous to the connection representation in
which the ``triad density" is represented by the functional derivative
 $\,^{\underline{*}}\underline{\vartheta}^{\alpha}:= \delta / \delta
{\buildrel\pm\over {\underline{A}_{\alpha}}} $ with respect to the
canonically conjugate Ashtekar variable
$\buildrel\pm\over{ \underline{A}_\alpha}$. This provides a mapping
from the Hamiltonian constraint of gravity with cosmological term
to the Chern--Simons 3--form [7] of the Ashtekar--Sen connection.} to
each other
via $\vartheta^{A}=c{\buildrel\circ\over \Gamma}{}^{\star A}=
{c\over 2}\eta^{ABC}
{\buildrel\circ\over \Gamma}_{BC}$, where $A,B,C =0,1,2$ (or $=1,2,3$ for
Euclidean signature). For
${\buildrel\circ\over \Gamma}_{AB}\rightarrow 0$, a
degenerate coframe (``triad") will occur in this model. By regarding
coframe and ``Lorentz"--rotational connection as part of the
{\it Cartan connection}
$${\buildrel =\over\Gamma}=
\vartheta^{A}\, P_{A} +{\buildrel\circ\over \Gamma}{}^{BC}\,
{\buildrel\circ\over L}_{BC}\, ,\eqno(\z)$$
Witten [8] could show that the 3D Hilbert--Einstein Lagrangian can be
absorbed in
a Chern--Simons term for (1.3), thus facilitating the proof of the
{\it finiteness} of the corresponding 3D quantum model.

Degenerate coframes, however, tend to jeopardize the coupling
of gravity to matter fields, as exemplified by Dirac or Rarita--Schwinger
fields. The basic reason is that the local frame $e_{\alpha}$, even if
it still exists, is not invertible any more; i.e. the relation
$e_{\alpha}\rfloor\vartheta^{\beta} =\delta_{\alpha}^{\beta}\>$, which
is needed in the formulation of the matter Lagrangian, would then
be lost. In this paper, we want to resolve this riddle by demonstrating
explicitly that $\vartheta^{\alpha}$ is only part of
the dimensionless translational gauge potential, if we use a
Yang--Mills type gauge approach to the affine group, which includes the
Poincar\'e group of elementary particle physics as subgroup. Thereby we
also clarify the subtle relationship between infinitesimal translational
gauge transformations and four
one--parameter subgroups of diffeomorphism of
spacetime, both regarded as acting {\it actively}.

\bigskip\goodbreak
\sectio{\bf The rigid affine group $A(n,R)$}

In the flat $n$--dimensional affine space $R^n$, the rigid
{\it affine group} $A(n,R):= R^{n}\semidirect$ $GL(n,R)$ acts as the
semidirect product of the group of n--dimensional translations and
n--dimensional general linear transformations. Thus it is a
generalization of the Poincar\'e group $P:= R^{4}\semidirect SO(1,3)$,
with the pseudo--orthogonal group $SO(1, n-1)$ being replaced by the
general linear group
$GL(n, R)$. In the following, we will work in a M\"obius
type representation [9, 10]. The $A(n,R)$ is that subgroup of
$GL(n+1,R)$
which leaves the n--dimensional hyperplane
${\buildrel =\over R}{}^{n}:=\left\{{\buildrel =\over x} =
\pmatrix{x\cr 1\cr}\in R^{n+1}\right\}$
invariant:
$$A(n,R) = \left\{ \pmatrix{\Lambda& \tau\cr 0& 1\cr}\in
GL(n+1,R)\left\vert \; \Lambda\in GL(n,R), \; \tau\in
R^n\right.\right\}.
\eqno(\z)$$
Thus we obtain ${\buildrel =\over x}' =\pmatrix{\Lambda x+\tau\cr 1\cr}$
as it is required for the action of the affine group on the flat affine
space.

The Lie algebra $a(n,R)$ consists of the generators
$P_{\gamma}$ representing local
n--dimen\-sional translations and the $L^{\alpha}{}_{\beta}$ which span the
Lie algebra $gl(n,R)$ of n--dimensional linear transformations.
Their commutation relations are:
$$[P_{\alpha}\, , P_{\beta}]= 0,
\eqno(\z)$$
$$[L^{\alpha}{}_{\beta}\, ,  P_{\gamma}] =
\delta^{\alpha}_{\gamma}\, P_{\beta}\, ,
 \eqno(\z)$$
and
$$[L^{\alpha}{}_{\beta},  L^{\gamma}{}_{\delta}] =
\delta^{\alpha}_{\delta}\, L^{\gamma}{}_{\beta} -
\delta^{\gamma}_{\beta}\, L^{\alpha}{}_{\delta}\, . \eqno(\z)$$
Observe that the physical dimensions of these generators are
$[L^{\alpha}{}_{\beta}\, ]=\hbar$ and $[P_{\alpha}\, ]=\hbar/{\rm
length}$.
If there is a spacetime metric $g=g_{\alpha\beta}\,
\vartheta^{\alpha}\otimes \vartheta^{\beta}$ with components
$g_{\alpha\beta}$ available for
lowering indices, the general linear group can be decomposed further:
$$L_{\alpha\beta} = \left(\;
{\buildrel\circ\over L}_{\alpha\beta} + L\quer _{\alpha\beta}
+ {1\over n} \; g_{\alpha\beta}\, {\cal D}\; \right)\, .
 \eqno(\z) $$
Here ${\buildrel\circ\over L}_{\alpha\beta}:=L_{[\alpha\beta]}$
are the generators of Lorentz transformations,
$L\quer _{\alpha\beta}:=L_{(\alpha\beta )}-(1/n)\,g_{\alpha\beta}\,
L^\gamma{}_\gamma$ represent shears,
whereas ${\cal D}:=L^{\gamma}{}_{\gamma}$ is the
generator of scale transformations.

\bigskip\goodbreak
\sectio{\bf The affine gauge approach}

In a matrix representation analogous to (2.1), we can write the affine
gauge group\myfoot{In a fibre bundle approach, one introduces first
the {\it bundle of affine frames}
$A(M):= P\Bigl(M^{n}, A(n,R), \pi, \delta \Bigr)$
where $\pi$ denotes the projection to the base manifold and
$\delta$ the (left) action of the structure group $A(n,R)$ on the bundle.
Active, affine gauge transformations are the vertical automorphisms
of $A(M)$. Similarly as the diffeomorphisms of the base manifold $M^n$,
they form the infinite--dimensional group
${\cal A}(n,\, R):=C^{\infty}\Bigl(A(M)\times_{Ad}\, A(n,\, R)\Bigr)$.
The group
${\cal G}{\cal L}(n,\, R):=C^{\infty}(A(M)\times_{Ad}\,GL(n,\, R)) $
of linear gauge transformations and the group
${\cal T}(n,\, R):=C^{\infty}(A(M)\times_{Ad}\, R^{n}) $
of local translations are subgroups of ${\cal A}(n,\, R)$.Taking the
cross--section in the associated bundle is abbreviated by
$C^{\infty}$ and $Ad$ denotes the
adjoint representation with respect to $GL(n,R)$. Due to its
construction, the group of local translations ${\cal T}(n,\, R)$
is {\it locally} isomorphic to the group of active
diffeomorphisms $Diff(n,R)$ of the manifold [11, 12]. The
infinite--dimensional group $Diff(n,R)$ contains the $n+n^2$ --dimensional
group $A(n,R)_{H}$ of {\it holonomic} affine transformations [13]
which are generated by the vector fields $P_{i}=\partial_{i}$
$:=\partial /\partial x^{i}$ and $L^{i}{}_{j} =x^{i}\partial_{j}$ as a
subgroup. Note that differentiable coordinate transformations, which
leave exterior forms invariant, are regarded as passive
diffeomorphisms.} as
$${\cal A}(n,R) = \left\{ \pmatrix{\Lambda(x)& \tau (x)\cr
                           0& 1\cr}\left\vert \;
              \Lambda(x)\in {\cal G}{\cal L}(n,R), \quad \tau (x)\in
{\cal T}(n,R)\right.\right\}\, ,\eqno(\z)$$
Following a Yang--Mills type gauge approach, we introduce
the {\it generalized affine} connection [14], cf. [15],
$${\buildrel\approx\over\Gamma} =\pmatrix{\Gamma^{(L)}& \Gamma^{(T)}\cr
                           0& 0\cr}
=\pmatrix{\Gamma^{(L)}_{\alpha}{}^{\beta}\,L^{\alpha}{}_{\beta}&
\Gamma^{(T)\alpha}\,P_{\alpha}\cr
                           0& 0\cr}\eqno(\z)$$
and require that it transforms inhomogeneously under
an affine gauge transformation:
$${\buildrel\approx\over\Gamma}\quad
{\buildrel {A^{-1}(x)}\over\longrightarrow}\quad
{\buildrel\approx\over\Gamma}'=
A^{-1}(x)\,{\buildrel\approx\over\Gamma}\, A(x) -
 A^{-1}(x)dA(x)\; ,\qquad A(x)\in {\cal A}(n,\, R).\eqno(\z)$$
Since we regard it as an {\it active} transformation, it is formed
with respect to the group element
$$A^{-1}(x)=\pmatrix{\Lambda^{-1}(x)& -\Lambda^{-1}(x)\tau (x)\cr
                           0& 1\cr}\eqno(\z)$$
which is inverse to $A(x)\in {\cal A}(n,R)$.
The corresponding affine curvature is given by
$${\buildrel\approx\over R}:= d{\buildrel\approx\over\Gamma} +
{\buildrel\approx\over\Gamma}\wedge {\buildrel\approx\over\Gamma} =
\pmatrix{d\Gamma^{(L)} +\Gamma^{(L)}\wedge\Gamma^{(L)} &
d\Gamma^{(T)}+ \Gamma^{(L)}\wedge \Gamma^{(T)}\cr
                           0& 0\cr}\, ,\eqno(\z)$$ and transforms
covariantly, i.e.  $${\buildrel\approx\over R}\quad{\buildrel
{A^{-1}(x)}\over\longrightarrow} \quad {\buildrel\approx\over R}'=
A^{-1}(x)\,{\buildrel\approx\over R}\, A(x) \eqno(\z)$$ under the
affine gauge group.\myfoot{Our matrix formalism, cf. [10, 16] and
references therein, is a spacetime generalization of the so-called
{\it motor calculus} of von Mises [17].} The covariant exterior
derivative ${\buildrel\approx\over D}:= d + {\buildrel\approx\over
\Gamma}$ acts on an affine p--form ${\buildrel\approx\over\Psi} =
\pmatrix{\Psi\cr 1\cr}$ as follows $${\buildrel\approx\over
D}\,{\buildrel\approx\over\Psi}= \pmatrix{D\Psi +\Gamma^{(T)}\cr
                   0\cr}\, .\eqno(\z)$$
Only by imposing the gauge $\Gamma^{(T)}=0$, one would recover the
covariant exterior derivative $D :=d + \Gamma^{(L)}$ with respect
to the linear connection.

After insertion of (3.2) and (3.4), the inhomogeneous transformation
law (3.3) splits into
$$\Gamma^{(L)}\quad{\buildrel {A^{-1}(x)}\over\longrightarrow}\quad
\Gamma^{(L)}{}'= \Lambda^{-1}(x)\,\Gamma^{(L)}\, \Lambda(x) -
 \Lambda^{-1}(x)d\Lambda(x)\; ,\eqno(\z)$$
and
$$\Gamma^{(T)}\quad{\buildrel {A^{-1}(x)}\over\longrightarrow}\quad
\Gamma^{(T)}{}'= \Lambda^{-1}(x)\, \Gamma^{(T)} -
 \Lambda^{-1}(x)D\tau (x)\; .\eqno(\z)$$
The local translations $\tau (x)$ automatically drop out in (3.8) due
to the 1--form structure of $\Gamma^{(T)}$. Thereby (3.8) aquires the
conventional transformation rule (with the exterior derivative $d$)
for a Yang--Mills--type connection for
${\cal G}{\cal L}(n,R)$, and thus we can identify
$\Gamma^{(L)}=\Gamma= \Gamma_{\alpha}{}^{\beta}\, L^{\alpha}{}_{\beta}$
with the linear connection. Due to the covariant exterior derivative term
$D\tau (x):= d\tau (x)- \Gamma^{(L)}\, \tau (x)$ in (3.9), the
translational part $\Gamma^{(T)}$ does not
transform as a covector as is required for the coframe
$\vartheta :=\vartheta^{\alpha}\, P_{\alpha}$, i.e. the 1--form with
values in the Lie algebra of $R^n$.

However, we may follow Trautman [18] and introduce a
vector--valued $0$--form
 ${\buildrel\approx\over\xi} = \pmatrix{\xi\cr 1\cr}=
\pmatrix{\xi^{\alpha}\, P_{\alpha}\cr 1\cr}$
 which transforms as ${\buildrel\approx\over\xi}'=A^{-1}(x)\,
{\buildrel\approx\over\xi}$, i.e.,
$$\xi\quad {\buildrel {A^{-1}(x)}\over\longrightarrow}\quad
\xi'=\Lambda^{-1}(x)\,\bigl (\xi- \tau (x)\bigr)\eqno(\z)$$
under an active affine gauge transformation.  Then
$$\vartheta :=\Gamma^{(T)} - D\xi \eqno(\z)$$
transforms as a vector--valued $1$--form under the ${\cal A}(n,R)$,
as required:
$$\vartheta\quad {\buildrel {A^{-1}(x)}\over\longrightarrow}\quad
\vartheta'=\Lambda^{-1}(x)\, \vartheta\, .\eqno(\z)$$

If $\Gamma^{(T)}$ vanished throughout the manifold, the vector field
$\xi$ would represent a four dimensional version of Cartan's
{\it generalized radius vector} [19]. The integrability condition is,
in this instance, given by the vanishing of
the translational part of the affine curvature (3.5), i.e.,
$R^{(T)}:= D\Gamma^{(T)}=(T^{\alpha} + R_{\beta}^{(L)\alpha}\,
\xi^{\beta}\,)P_{\alpha} =0$, which, for teleparallelism
models with $ R_{\alpha}^{(L)\beta}=0$, would imply vanishing
torsion.\myfoot{Otherwise there also exists, for example, the nontrivial
solution
$T^{\alpha}=c\eta^{\alpha\beta}\xi_{\beta}$ and
$R^{\alpha\beta}=c\eta^{\alpha\beta}$, with the dimensionful constant c.}

\bigskip\goodbreak
\sectio{\bf Reduction to a Cartan connection}

Our key relation (3.11) in components takes the form
$$\Gamma_{i}^{(T)\alpha} = e_{i}{}^{\alpha} +
D_{i}\xi^{\alpha}\,,\eqno(\z)$$
which for $D_{i}\xi^{\alpha}=-\delta_{i}^{\alpha}$ makes contact with
an approach of Hayashi et al. [20]. In a recent paper [21] on the
Poincar\'e gauge approach,  our $\xi^{\alpha}$ are kinematically
interpreted as ``Poincar\'e coordinates"; note that in Eq. (2.14) of
that paper [21], vielbein and translational connection are identified
opposite to our notation. Observe also that we do not have to put the
``Poincar\'e coordinates" $\xi^{\alpha}$ to zero, in order to obtain
the gauge affine (!) transformation law (3.12) of the coframe. The reason
is that the local translations are now ``hidden" in the invariant
transformation behavior of the exterior 1--form  $\vartheta$ under
(passive) diffeomorphisms. Note also that in our approach, in
contradistinction to that of Sexl and
Urbantke [22, p. 381], we do not need to break the affine gauge group
kinematically via $D\tau (x)=0$. An attempt to motivate the translational
connection (4.1) from the theory of dislocations can be found
in Ref. [23], whereas Hennig and Nitsch [24] provide an
explanation in terms of jet bundles.

Since $\xi=\xi^{\alpha}P_{\alpha}$ aquires its
values in the ``orbit" (coset space) $A(n,R)/GL(n,R)\approx R^n$,
it can be regarded as an affine vector field (or
``generalized Higgs field" according to
Trautman [25]) which ``hides" the action of the
local translational ``symmetry" ${\cal T}(n,R)$.
 If we required [26] the condition
$$D\xi=0\, ,\eqno(\z)$$
the translational connection $\Gamma^{(T)}$ would, together with the
coframe $\vartheta$, be ``soldered'' to the spacetime manifold,
cf. [27], and the translational part of the local affine group
would be ``spontaneously broken", cf. [28]. The stronger constraint of
a ``zero section" vector field  $\xi= 0$ would
reduce the generalized affine connection
${\buildrel\approx\over\Gamma}$ on the affine bundle $A(M)$
to the {\it Cartan connection} [9]
$${\buildrel =\over\Gamma} =\pmatrix{\Gamma&  \vartheta\cr
                           0& 0\cr} \eqno(\z)$$
on the bundle $L(M)$ of linear frames. Due to (3.12), this is not
anymore a connection in the usual sense. However, thereby
we would recover the familiar (metric--)\-affine geometrical
arena [29] with nonmetricity, torsion, and
curvature, as is summarized in the following table:

$$\vbox{\offinterlineskip
\hrule
\halign{&\vrule#&\strut\quad\hfil#\quad\hfil\cr
height2pt&\omit&&\omit&&\omit&\cr
& {\bf potential } && {\bf field strength } && {\bf Bianchi identity } &\cr
height2pt&\omit&&\omit&&\omit&\cr
\noalign{\hrule}
height2pt&\omit&&\omit&&\omit&\cr
& metric $g^{\alpha \beta }$ && $Q^{\alpha \beta }= Dg^{\alpha \beta }$ &&
$DQ^{\alpha \beta }= 2 R_\mu{}^{(\alpha}\, g^{\beta )\mu}$ &\cr
& coframe  $\vartheta ^\alpha $  &&
$T^\alpha = D\vartheta ^\alpha $ && $DT^\alpha = R_\mu {}^\alpha \wedge
\vartheta ^\mu $ &\cr
& connection $ \Gamma _\alpha {}^\beta $ && $ R_\alpha {}^\beta
= d \Gamma _\alpha {}^\beta + \Gamma _\mu {}^\beta \wedge
\Gamma _\alpha {}^\mu $ && $D  R_\alpha {}^\beta =0 $ &\cr
height2pt&\omit&&\omit&&\omit&\cr}
\hrule}$$
\bigskip

\bigskip\goodbreak
\sectio{\bf Affine gauge transformations versus active diffeomorphisms}

The affine gauge transformations in (3.3) are finite transformations.
If we expand them up to first order according to
$$\Lambda (x)= 1 + \omega_{\alpha}{}^{\beta}\,
 L^{\alpha}{}_{\beta} + \cdots\, ,\eqno(\z)$$
$$\tau (x) = 0 + \varepsilon^{\alpha}\, P_{\alpha}
+ \cdots\, ,\eqno(\z)$$
we obtain from (3.8) and (3.9), respectively,
$$\delta_{A^{-1}}\Gamma^{(L)}= -(D\omega_{\alpha}{}^{\beta})\,
L^{\alpha}{}_{\beta} + \cdots\, ,\eqno(\z)$$
$$\delta_{A^{-1}}\Gamma^{(T)}= -(D\varepsilon_{\alpha} +
\omega_{\beta}{}^{\alpha}\, \Gamma^{(T)\beta})\,P_{\alpha}
 + \cdots\, .\eqno(\z)$$
(For the ``product" of Lie generators we use the Lie brackets of Sect.~2,
since we work in the adjoint representation.) It is gratifying to note
that the leading exterior covariant derivatives reveal,
in particular, that the translational connection $\Gamma^{(T)}$ is really
the ``compensating" field for infinitesimal local translations $\varepsilon$
in the Yang--Mills sense.

Let us compare this result with the ``diffeomorphism" approach,
which was orginally developed for the Poincar\'e subgroup of the $A(n,R)$:
In essence, the translational part of the transformation
$$\Pi= 1-\varepsilon -\omega = 1 - \varepsilon^{\alpha}\, P_{\alpha} -
\omega_{\alpha}{}^{\beta}\, L^{\alpha}{}_{\beta},\eqno(\z)$$
is embedded as an $n$--parameter subgroup of the
infinite--dimensional group of {\it active} diffeomorphisms of
spacetime.\myfoot{The minus sign is in accordance with
our earlier conventions for active trans\-for\-mations.} In order
to calculate the effect on the linear conection and the coframe, one
has to consider the action [30] of the Lie derivative
${\cal L}_{(-\varepsilon)}$ with respect to the vector
field $(-\varepsilon )$ together with an infinitesimal frame
rotation parametrized by $(-\omega )$. Since ${\cal L}_{(-\varepsilon)}=
\ell_{(-\varepsilon)}:= -(\varepsilon\rfloor d +d\varepsilon\rfloor) $
holds for geometrical objects which are invariant under changes of
the basis, a straightforward calculation yields
$$({\cal L}_{(-\varepsilon)}+\delta_{(-\omega)})\Gamma=
-\Bigl[D(\omega_{\alpha}{}^{\beta} +
\varepsilon\rfloor \Gamma_{\alpha}{}^{\beta})
+\varepsilon\rfloor R_{\alpha}{}^{\beta}\Bigr]\,L^{\alpha}{}_{\beta}
\, ,  \eqno(\z)$$
and
$$({\cal L}_{(-\varepsilon)}+\delta_{(-\omega)})\vartheta=
-\Bigl[D\varepsilon^{\alpha} -
(\omega_{\beta}{}^{\alpha} +
\varepsilon\rfloor\Gamma_{\beta}{}^{\alpha})\vartheta^{\beta}
+\varepsilon\rfloor T^{\alpha}\Bigr]\,P_{\alpha}
\, .  \eqno(\z)$$

The `annoying' linear connection term in (5.6) and (5.7) can be
dismissed by going
over to the {\it parallel transport} version of Hehl et al. [2] and
Ne'eman [31] in which, instead of $P_{\alpha}$, the covariant derivative
components $D_{\alpha}:=e_{\alpha}\rfloor D$ are adopted as generators
of local translations: Then the infinitesimal transformations read
$$\tilde\Pi=  1 - \varepsilon^{\alpha}\, D_{\alpha} -
\omega_{\alpha}{}^{\beta}\, L^{\alpha}{}_{\beta}=
\Pi +\varepsilon\rfloor \Gamma_{\alpha}{}^{\beta}\,
L^{\alpha}{}_{\beta}\, .\eqno(\z)$$

Since this amounts to a redefinition
$\tilde\omega := \omega -\varepsilon\rfloor \Gamma_{\alpha}{}^{\beta}\,
L^{\alpha}{}_{\beta}$ of the parameters of the infinitesimal
linear transformation, we can simply read off, from (5.6) and (5.7),
the new results
$$({\cal L}_{(-\varepsilon)}+\delta_{(-\tilde\omega)})\Gamma=
-\Bigl[D\omega_{\alpha}{}^{\beta} +
\varepsilon\rfloor R_{\alpha}{}^{\beta}\Bigr]\,L^{\alpha}{}_{\beta}
\, ,  \eqno(\z)$$
and
$$({\cal L}_{(-\varepsilon)}+\delta_{(-\tilde\omega)})\vartheta=
-\Bigl[D\varepsilon^{\alpha} -
\omega_{\beta}{}^{\alpha}\vartheta^{\beta}
+\varepsilon\rfloor T^{\alpha}\Bigr]\,P_{\alpha}\, .  \eqno(\z)$$

In this {\it parallel transport} version, the leading
covariant derivative pieces are the same as in the affine gauge
approach. In particular, the ``hidden" translational piece in
the {\it affine} transformation (3.12) of the coframe gets thereby
``uncovered" in (5.10). Is it in the end ``...somewhat a matter of
taste...", as Nester [32] has put it, whether or not one prefers the
parallel transport interpretation of translations over the affine
gauge approach? One could argue that the Pauli--type curvature and
torsion terms in the infinitesimal transformations (5.9) and
(5.10) violate the spirit of the principle of {\it ``minimal coupling"}, a
cornerstone of a conventional Yang--Mills type gauge approach. These terms
also show up in the commutation relation
$$[D_{\alpha}\, ,D_{\beta}]= -T_{\alpha\beta}{}^{\gamma} D_{\gamma} +
\,R_{\alpha\beta\mu}{}^{\nu}\, L^{\mu}{}_{\nu}
\,\eqno(\z)$$
for the operator $D_{\alpha}$ of parallel--transport, if applied to
a 0--form. Due to the torsion and curvature terms on the right--hand side,
a {\it softening} [33] of the Lie algebra structure cannot be
avoided in such a diffeomorphism--type approach. Using the covariant
derivatives (or Lie derivatives) has the advantage of being physically
meaningful as a parallel transport, as explained in Ref. [2],
once we put up a frame, and, in a corresponding first order approach, these
`non--minimal' structures do {\it not}
touch the explicit form of the Lagrangian.
However,  they are algebraically less useful because (5.11) is not
a Lie algebra any more. Moreover, as we will show below,
the affine gauge approach lends itself to an important resolution of
degeneracy problems in quantum gravity.

\bigskip\goodbreak
\sectio{\bf Affine gauge approach to quantum gravity with topology
change}

Now we may return to the question of the proper meaning and range
of validity of the functional integral (1.2) in quantum gravity.
The lesson learnt from  our affine gauge approach is that, instead of
the coframe,  a summation over the true translational connection
$\Gamma^{(T)}$ is more akin to a quantum Yang--Mills type approach.
Moreover, it would cause no problems if the functional integration
didn't go only through $\Gamma^{(L)}=0$ as it would do in Yang--Mills
theory, but through  $\Gamma^{(T)}=0$
as well, keeping the coframe $\vartheta$, by definition,
non--degenerate. In effect, we may now consider the functional integral
$$\int  {\cal D}{\buildrel\approx\over\Gamma}\;
exp\bigr( i\int_{M}V_{\rm EC}\bigl),\eqno(\z)$$
where summation is understood over the {\it generalized affine} connection.

Due to (3.8) and (3.12), the Ein\-stein--Cartan Lagrangian
is gauge invariant also with respect to the full, but ``hidden", affine
gauge group ${\cal A}(4, R)$. In order to study the possibility of a
degenerate or even vanishing translational connection, it is
instructive to insert the
representation (3.11) of the coframe into the Ein\-stein--Cartan
Lagrangian (1.1) amended by a cosmological term
$(-\Lambda_{\rm cos}/\ell^{2})\eta$. In the
gauge $\Gamma^{(T)}\rightarrow 0$, we find then
the following truncated expression:
$$\eqalign{V_{\rm ECC}
&=- {1\over{ 4\ell^{2}}}{\buildrel\circ\over R}{}^{\alpha\beta}
\wedge (\Gamma^{(T)\gamma} -D\xi^{\gamma})\wedge
 (\Gamma^{(T)\delta} -D\xi^{\delta})
\eta_{\alpha\beta\gamma\delta}\, \cr
\qquad &\qquad \cr
&- {\Lambda_{\rm cos}\over{ 4!\ell^{2}}}
(\Gamma^{(T)\alpha} -D\xi^{\alpha})\wedge
(\Gamma^{(T)\beta} -D\xi^{\beta})\wedge
(\Gamma^{(T)\gamma} -D\xi^{\gamma})\wedge
 (\Gamma^{(T)\delta} -D\xi^{\delta})
\eta_{\alpha\beta\gamma\delta}\, \cr
\qquad &\qquad \cr
&\longrightarrow\quad {1\over{ 4\ell^{2}}}
{\buildrel\circ\over R}{}^{\alpha\beta}\wedge \xi^{\gamma}
{\buildrel\circ\over R}_{\mu}{}^{\delta}\, \xi^{\mu}
\eta_{\alpha\beta\gamma\delta}\,+
{\Lambda_{\rm cos}\over{ 4!\ell^{2}}}
\xi^{\alpha} D\xi^{\beta}\wedge D\xi^{\gamma}\wedge
{\buildrel\circ\over R}_{\mu}{}^{\delta}\, \xi^{\mu}
\eta_{\alpha\beta\gamma\delta}\, \cr
\qquad &\qquad \cr
&\qquad -
{1\over{ 4\ell^{2}}} d\Bigl[({\buildrel\circ\over R}{}^{\alpha\beta}
\wedge \xi^{\gamma}\,D\xi^{\delta}+
{\Lambda_{\rm cos}\over{ 3!\ell^{2}}}
\xi^{\alpha} D\xi^{\beta}\wedge D\xi^{\gamma}\wedge
D\xi^{\delta})
\eta_{\alpha\beta\gamma\delta}\Bigr]\, .\cr} \eqno(\z)$$
In order to separate off the boundary term, we employed the Bianchi
identity $D{\buildrel\circ\over R}{}^{\alpha\beta}=0$ for the
Riemann--Cartan curvature, which is valid in the Riemann--Cartan framework
of (6.2) with vanishing nonmetricity.

Thus the occurrence of a vanishing translational connection in the
functional integral (6.1) is without harm, since a quadratic and
linear curvature Lagrangian remains for the linear connection. In case
that $\Lambda_{\rm cos}=0$, except for the dimensional coupling
constant, the truncated expression (6.2) resembles the
Stephenson--Kilmister--Yang (SKY) Lagrangian [34, 35, 36], which is
known to be perturbatively renormalizable [37, 38]. Moreover, for the
constant vacuum condensate $\xi^{\hat 0}_{<>} =H \neq 0$ and zero
otherwise, the Lagrangian (6.2) reduces ``spontaneously" to the
Euclidean 3D topological gravity model $$V_{{\rm ECC} <>} = {1\over{
4\ell^2}}\Bigl[\eta_{ABC}\, {\buildrel\circ\over R}^{AB}\wedge
{\buildrel\circ\over R}_{\hat 0}{}^{C} H^{2} + {\Lambda_{\rm
cos}\over{ 3!}}\,\eta_{ABC}\, {\buildrel\circ\over\Gamma}_{\hat
0}{}^{A}\wedge {\buildrel\circ\over\Gamma}_{\hat 0}{}^{B}\wedge
{\buildrel\circ\over\Gamma}_{\hat 0}{}^{C}\, H^{4}\Bigr]\, \eqno(\z)$$
in which the dreibein can be identified with part of the Lorentz
connection via $\vartheta^{A}= {\buildrel\circ\over\Gamma}_{\hat
0}{}^{A}$ and
 $T^{A}={\buildrel\circ\over R}_{\hat 0}{}^{A}$. Other directions of
symmetry breaking, such as
 $\xi^{\hat 3}_{<>} =H \neq 0$, lead to Minkowskian models of
3D gravity [39], cf. [40]. For a time--dependent $H=H(t)$, one would
obtain, in addition to the curvature dependent $H^{2} + H^{4}$
potential in (6.3), a kinetic term for $H$ and could analyse, following
Giddings [41],  the {\it
instability} of the $\xi^{\alpha}=0$ solution. In the
gauge $\Gamma^{L}=0$, we thereby obtain a means to analyse
the instability not only for the diffeomorphism invariant
solution $\vartheta =0$  but also for the
true translation invariant solution $\Gamma^{T}=0$. It remains to be
seen, if also the signature of the physical spacetime has a
dynamical origin in such a framework, as is suggested by Greensite [42],
cf. [43].

In quantum gravity, a vanishing translational connection
$\Gamma^{(T)}$ may be accompanied by a {\it topology change}
of the underlying spacetime manifold. The rich spectrum of possible
topological structures in quantum ``geometrodynamics" has been outlined in
Ref.~[44]. Our affine gauge approach exactly leads us to the more
detailed mechanism devised by Horowitz [3]. According to (4.11), the
vanishing of the translational connection converts the coframe
components into the form $\vartheta^{\alpha} = - D\xi^{\alpha}$.
At each of the boundaries $\partial M^{4} \approx
S^{3} \bigcup S^{\prime 3}$ of the topology changing spacetime manifold
$M^{4}$ we may also adopt the gauge $\Gamma^{(L)}=0$ for the linear
(or Lorentz) connection. In the vicinity of that boundary,
$$\vartheta^{\alpha} = - d\xi^{\alpha}\qquad {\rm and}\qquad
g= o_{\alpha\beta}\, d\xi^{\alpha}\otimes d\xi^{\beta} \eqno(\z)$$
represent the (flat !) Minkowskian spacetime in terms of the
4--dimensional Cartesian coordinate system $\{ \xi^{\alpha}\}$.
Obviously, this solves also the
vacuum Einstein--Cartan theory with zero action. For generic
$\xi^{\alpha}$, the (inverse) tetrad components $e_{j}{}^{\beta} =-
\partial_{j}\,\xi^{\beta}$ will be {\it non--degenerate} almost
everywhere. In order to accomplish a spatial topology change, we may
choose $\xi^{\hat {0}}$ to be the ``height function" of
{\it Morse theory}. Then, $d\xi^{\hat {0}}$ is a timelike
covector and we can apply
the ``trouser world" construction as is decribed, for example, by
Konstantinov and Melnikov in Fig. 1 of Ref. [45].

We summarize: The advantage of the affine gauge approach is that the
coframe, even for vanishing translational connection, remains
non-degenerate almost everywhere.

\bigskip
\centerline{\bf  Acknowledgments}

We would like to thank J\"org Hennig (Clausthal) and Yuri Obukhov
(Cologne/ Moscow) for constructive comments on a preliminary version of
this paper.

\bigskip\goodbreak
\centerline {\bf References}

\newref
[1] R.P. Feynman: {\it Lectures on Gravitation.} Lecture notes by
F.B. Morinigo and W.G. Wagner (California Institute of Technology,
Pasadena, California 1962/63).
\newref
[2] F. Hehl, P. v. Heyde, G.D. Kerlick and J.M. Nester, {\it Rev.
Mod. Phys.} {\bf 48}, 393 (1976); see also
F.W.  Hehl: ``Four lectures on Poincar\'e gauge theory",
in: Proceedings of the 6th Course of the School of Cosmology and
Gravitation on {\it Spin, Torsion, Rotation,
and Supergravity}, held at Erice, Italy, May 1979, P. G. Bergmann,
V. de Sabbata, eds. (Plenum, New York 1980), p.5.

\newref
[3] G.T. Horowitz, Class. Quantum Grav. {\bf 8} (1991) 587.
\newref
[4] C.J. Isham: ``Conceptual and geometrical problems in
quantum gravity", in: {\it Recent Aspects of Quantum Fields},
Proceedings of the XXX$^{\hbox{th}}$ Int. Univer\-sit\"ats\-wo\-chen
f\"ur Kernphysik,
Schladming, Austria, February and March 1991, H. Mitter and H. Gausterer,
eds., Lecture Notes in Physics, Vol. 396 (Springer, Berlin 1991), p.123.
\newref
[5] E.W. Mielke, Ann. Phys. (N.Y.) {\bf 219} (1992) 78.
\newref
[6] P. Baekler, E.W. Mielke, and F.W. Hehl, Nuovo Cimento {\bf 107B}
(1992) 91.
\newref
[7] B. Br\"ugmann, R. Gambini, and J. Pullin, Nucl. Phys.
{\bf B 385} (1992) 587.
\newref
[8] E. Witten, Nucl. Phys. {\bf B311} (1988/89) 46.
\newref
[9] S. Kobayashi: {\it Transformation Groups in Differential Geometry}
(Springer, New York 1972).
\newref
[10] E.W. Mielke: {\it Geometrodynamics of Gauge Fields} --- On
the geometry of Yang-Mills and gravitational gauge
theories (Akademie-Verlag, Berlin 1987).
\newref
[11]  V.I. Ogievetsky, Lett. Nuovo Cimento
{\bf 8} (1973) 988.
\newref
[12] S. Sternberg, Ann. Phys. (N.Y.) {\bf 162} (1985) 85.
\newref
[13] P.G.  Bergmann and A.B. Komar, J. Math. Phys. {\bf 26} (1985) 2030.
\newref
[14] S. Kobayashi and K. Nomizu: {\it Foundations of Differential
Geometry}, Vol. I (Interscience Publ., New York 1963).
\newref
[15] L.K. Norris, R.O. Fulp, and W.R. Davies,  Phys. Lett.
{\bf 79A} (1980) 278.
\newref
[16] K.L.\ Malyshev and A.E.\ Romanov: ``Aspects of $T(3)\semidirect
SO(3)$-gauge theory of dislocations and disclinations I and II", in
Russian, Leningrad preprints LOMI P-2-90 and P-5-90 (1990).
\newref
[17] R.\ von Mises: ``Motorrechnung, ein neues Hilfsmittel der
Mechanik'', ZAMM {\bf 4} (1924) 155.
\newref
[18] A. Trautman: ``On the structure of the
Einstein--Cartan equations'', in: {\it Differential
Geometry}, Symposia Matematica
Vol.12  (Academic Press, London 1973) p.139.
\newref
[19] E. Cartan: {\it On a manifold with an affine connection
and the theory of general relativity} (Bibliopolis, Napoli 1986).
\newref
[20] K. Hayashi and T. Nakano, Progr. Theor. Phys. {\bf 38} (1967)
491; K. Hayashi and T. Shirafuji, Progr. Theor. Phys. {\bf 64} (1980)
866; {\bf 80} (1988) 711.
\newref
[21] G. Grignani and G. Nardelli, Phys. Rev. {\bf 45} (1992) 2719.
\newref
[22] R.U. Sexl and H.K. Urbantke: {\it Gravitation und Kosmologie},
2nd edition (Bibliographisches Institut, Mannheim 1983).
\newref
[23] G. Sardanashvily and M. Gogbershvily, Mod. Phys. Lett.
{\bf A2} (1987) 609.
\newref
[24] J. Hennig and J. Nitsch, Gen. Rel. Grav. {\bf 13} (1981) 947.
\newref
[25] A. Trautman,  Czech. J. Phys. {\bf B29} (1979) 107.
\newref
[26] K.A. Pilch, Lett. Math. Phys. {\bf 4} (1980) 49.
\newref
[27] P.K. Smrz, J. Math. Phys. {\bf 28} (1987) 2824.
\newref
[28] L. O'Raiffeartaigh: ``Some hidden aspects of hidden symmetry'',
in: {\it Differential Geometry, Group Representations, and
Quantization}, Lecture Notes in Physics, Vol. 379, J.D. Hennig,
W. L\"ucke, and J. Tolar, eds. (Springer, Berlin 1991), p. 99.
\newref
[29] F.W. Hehl, J.D. McCrea, E.W. Mielke, and Y. Ne'eman,
 Found. Phys. {\bf 19}, 1075 (1989); {\it Phys. Rep.} (to be published);
J.D. McCrea, {\it Class. Quantum Grav.} {\bf 9}, 553 (1992).
\newref
[30] R.D. Hecht, F.W. Hehl, J.D. McCrea, E.W. Mielke, and Y. Ne'eman:
``Improved energy--momentum  currents in metric--affine spacetime",
Phys. Lett. {\bf A} (in print 1992).
\newref
[31] Y.\ Ne'eman, in: {\it Differential Geometrical Methods in
Mathematical Physics}, K. Bleuler, H. R. Petry, and A. Reetz, eds.
Lecture Notes in
Mathematics, Vol. 676 (Springer-Verlag, Berlin 1978), p.189.
\newref
[32] J. M. Nester, in: {\it An Introduction to Kaluza-Klein Theories},
Workshop Chalk River/Deep River, Ontario 11-16 Aug. 1983,
H. C. Lee, eds. (World Scientific, Singapore 1984), p.83.
\newref
[33] T.W.B. Kibble and K.S. Stelle: ``Gauge theories of gravity and
Supergravity", in: {\it Progress in Quantum Field Theory: Festschrift for
Umezawa.} H. Ezawa and
S. Kamefuchi, eds. (Elsevier Science Publ. B.V. 1986), p. 57.
\newref
[34] G. Stephenson: Nuovo Cimento {\bf 9} (1958) 263.
\newref
[35] C.W. Kilmister and D.J. Newman, Proc. Cambridge Phil. Soc.
(Math. Phys. Sc.) {\bf 57} (1961) 851.
\newref
[36] C.N. Yang,  Phys. Rev. Lett. {\bf 33} (1974) 445.
\newref
[37] K.S. Stelle,  Phys. Rev. {\bf D16} (1977) 953.
\newref
[38] C.--Y. Lee and Y. Ne'eman,  Phys. Lett. {\bf B242} (1990) 59;
 C.--Y. Lee, Class. Quantum Grav. {\bf 9} (1992) 2001.
\newref
[39] T.T. Burwick, A.H. Chamseddine, and K.A. Meissner, {\it Phys. Lett.}
{\bf B284} (1992) 11.
\newref
[40] G. Grignani and G. Nardelli: ``Chern--Simons gravity from
$3+1$ dimensional gravity", preprint DFUPG--57--92.
\newref
[41] S.B. Giddings, Phys. Lett. {\bf B268} (1991) 17.
\newref
[42] J. Greensite, ``Dynamical origin of the Lorentz\-ian signa\-ture of
spacetime", pre\-print NBI--HE--92--59.
\newref
[43] A.D.\ Sakharov, {\it Sov.\ Phys.\ JETP} {\bf 60} (1985) 214.
\newref
[44] E.W. Mielke: Gen.\ Rel.\ Grav. {\bf 8} (1977) 175.
\newref
[45] M.Yu. Konstantinov and V.N. Melnikov: Class. Quantum Grav. {\bf
3} (1986) 401.
\bigskip\centerline{---------------}
\bye